\documentclass[preprint]{sig-alternate}

\title{A Local Mean Field Analysis \\ of Security Investments in Networks}

\numberofauthors{2}
\author{
\alignauthor
Marc Lelarge\\
       \affaddr{INRIA-ENS}\\
      \affaddr{Paris, France}\\
       \affaddr{marc.lelarge@ens.fr}
\alignauthor
Jean Bolot\\
       \affaddr{Sprint}\\
       \affaddr{California, USA}\\
       \affaddr{bolot@sprint.com}
}

\usepackage{psfrag, epsfig, amsmath,amssymb}
\def\EE{\mathbb{E}}
\def\PP{\mathbb{P}}
\def\NN{\mathbb{N}}

\def\ind{{\rm 1\hspace{-0.90ex}1}}

\def\o{{\O}}

\def\=d{\stackrel{d}{=}}

\def\gen{{\rm{gen}}}

\newtheorem{proposition}{Proposition}

\newtheorem{remark}{Remark}

\def\ind{{\rm 1\hspace{-0.90ex}1}}

\begin{document}
\toappear{to appear in NetEcon'08 \cite{lbp08}. This version includes proofs (given in Appendix) of results stated in \cite{lbp08}.}
\maketitle

\begin{abstract}
Getting agents in the Internet, and in networks in general,
to invest in and deploy security features and protocols is a challenge,
in particular because of economic reasons
arising from the presence of network externalities.
Our goal in this paper is to model and investigate
the impact of such externalities on security investments in a network.

Specifically, we study a network of interconnected agents subject to
epidemic risks such as viruses and worms where agents can
decide whether or not to invest some amount to deploy security
solutions. We consider both cases when the security solutions are strong (they
perfectly protect the agents deploying them) and when they are weak.
We make three contributions in the paper. First,
we introduce a general model which combines an epidemic propagation model with an economic
model for agents which captures network effects and externalities.
Second, borrowing ideas and techniques used in statistical
physics, we introduce a Local Mean Field (LMF) model, which extends the standard
mean-field approximation to take into account the correlation structure on
local neighborhoods. Third, we solve the LMF model in a network
with externalities, and we derive analytic solutions for
sparse random graphs of agents, for which we obtain asymptotic results.
We find known phenomena  such as free riders and tipping points. We also observe counter-intuitive phenomena, such as increasing the quality of the security technology
 can result in a decreased adoption of that technology in the network.
 In general, we find that both situations with
strong and weak protection exhibit externalities and that the equilibrium is not socially optimal - therefore there is a market failure. Insurance is one mechanism to address this market failure.
In related work, we have shown that insurance is a very effective mechanism \cite{bollel08, bollel08b} and argue that using insurance would increase the security in a network such as the Internet.

\end{abstract}

\keywords{Security, Game Theory, Epidemics, Economics, Price of Anarchy, Tipping, Free rider problem.} 

\section{Introduction}

Users and computers in the Internet face a wide range of security risks. Of particular concern,
are {\em epidemic risks}, such as those propagated by worms and viruses. Epidemic risks
depend on the behavior of
other entities in the network, such as whether or not those entities invest in
security solutions to minimize their likelihood of being infected.
Our goal in this paper is to analyze the strategic behavior of agents facing such
epidemic risks.

The propagation of worms and viruses \cite{zou02,ganesh05}, but also many other phenomena
in the Internet such as the propagation
 of alerts and patches \cite{vojnovic05} or of routing updates \cite{coffman02}, can be modeled using epidemic spreads through a network.
As a result, there is now a vast body of literature on epidemic spreads over a network topology from an initial set of infected nodes to
susceptible nodes. However, much of that work has focused on modeling and understanding
the propagation of the epidemics proper,  without considering
the impact of network effects and externalities.

Recent work which did model such effects
has been limited to the simple case of two agents, i.e. a two-node network.
For example, reference \cite{kunreuther03} proposes a
parametric game-theoretic model for such a situation. In the model, agents decide whether or not to invest in security and agents face a risk
of infection which depends on the state of other agents. The authors show the existence of two Nash equilibria (all agents invest or none invests), and
suggest that taxation or insurance would be ways to provide incentives for agents to invest (and therefore reach the "good" Nash equilibrium).
However, their approach does not scale to the case of $N$ agents, and it does not handle various network topologies
connecting those agents. Our work  addresses precisely those limitations.

The rest of the paper is organized as follows.
In Section \ref{sec:model1}, we describe our  model for epidemic risks with network effects and externalities.
In Section \ref{sec:lmf}, we introduce our Local Mean Field Model (LMF) and state asymptotic
results that can be obtained with LMF.
In Section \ref{sec:netext}, we use the  LMF model to examine both cases when agents invest
 in strong security solutions (which perfectly protect the agents deploying them against propagated risks) and in weak solutions. We find known phenomena  such as free riders and tipping points \cite{bollel08b}. We also observe counter-intuitive phenomena, such as increasing the quality of the security technology
 can result in a decreased adoption of that technology in the network.
In Section \ref{sec:discussion}, we discuss our results and conclude the paper.

\section{A model for epidemic risks and network effects}
\label{sec:model1}

\subsection{Economic model for the agents}\label{a:eco}
We model agents using the classical expected utility model, where agents  attempt to maximize a utility function $u$. We assume that agents are rational and that they are risk averse, i.e. their utility function is concave
(see Proposition 2.1 in \cite{gollier01}). Risk averse agents dislike
mean-preserving spreads in the distribution of their final wealth.

We denote by $w$ the initial wealth of the agent.
The {\it risk premium} $\pi$ is the maximum amount of money that one is ready to pay to escape a pure risk $X$, where a pure risk $X$ is a random
variable such that $\EE[X]=0$. The risk premium corresponds to an amount of money paid (thus decreasing the wealth of the agent from $w$ to $w - \pi$) which
covers the risk; hence, $\pi$ is given by the following equation:
$u[w - \pi] = \EE[u[w+X]]$.


Each agent faces a potential loss $\ell$, which we take in this paper to be a fixed (non-random) value. We denote by $p$ the probability of loss
or damage. There are two possible final
states for the agent: a good state, in which the final wealth of the agent
is equal to his initial wealth $w$, and a bad state in which the final wealth is $w-\ell$.
If the probability of loss is $p>0$, the risk is clearly not a pure risk. The amount of money $m$ the agent is ready to invest to
escape the risk is given by the equation:
$p u[w-\ell]+ (1-p)u[w] = u[w-m].$
We clearly
have $m>p\ell$ thanks to the concavity of $u$.
We can actually relate $m$ to the risk premium defined above:
\begin{eqnarray*}
m = p\ell+\pi[p].
\end{eqnarray*}

An agent can invest some amount in self-protection, which in practice would reflect
an investment in antivirus or anomaly detection solutions.
If an agent decides
to invest in self-protection, we say that the agent is in state $S$
(as in Safe or Secure).
If the agent decides not to invest in self-protection, it is
in state $N$ (Not safe).
If the agent does not invest, its probability of loss is $p^N$.
If it does invest,
for an amount which we assume is a fixed amount  $c$, then its loss
probability is reduced and equal to $p^S < p^N$.

In state $N$, the expected utility of the agent is $p^Nu[w-\ell]+(1-p^N)u[w]$;
in state $S$, the expected utility is
$p^Su[w-\ell-c]+(1-p^S)u[w-c]$. Using the definition of risk
premium, we see that these quantities are
equal to $u[w-p^N\ell-\pi[p^N]]$ and $u[w-c-p^S\ell-\pi[p^S]]$,
respectively. Therefore, the optimal strategy is for the agent to
invest in self-protection only if the cost for self-protection is less
than the threshold
\begin{eqnarray}
\label{eq:optic}c < (p^N-p^S)\ell +\pi[p^N]-\pi[p^S].
\end{eqnarray}

\subsection{Epidemic model}\label{sec:stepid}

We describe now our model for the epidemic risk.
Agents are represented by vertices of a graph.
We assume that an agent in state $S$ has a probability $p^-$ of direct
loss and an agent in state $N$ has a probability $p^+$ of direct loss
with $p^+\geq p^-$.
Then any infected agent contaminates neighbors independently of each
others with probability $q^-$ if the neighbor is in state $S$ and
$q^+$ if the neighbor is in state $N$, with $q^+\geq q^-$.

Special cases of this model are examined in \cite{lelbol08}, where $q^+=q^-$,
and in \cite{msw06}, where agents in state $S$ are completely secure and cannot be
infected, i.e. $p^-=q^-=0$.

Let $G = (V, E)$ be a graph on a countable vertex set $V$. Agents
are represented by vertices of the graph. For
$i,j\in V$, we write $i\sim j$ if $(i,j)\in E$ and we say that agents
$i$ and $j$ are neighbors.
The state of agent $i$ is represented by
$X_i$; agent $i$ is infected (respectively healthy) iff $X_i=1$
(respectively $X_i=0$).

We now describe the fundamental recursion satisfied by the vector $X$.
We first introduce the following sequences of independent identically
distributed (i.i.d.) random variables (r.v.):
\begin{itemize}
\item $(A^S,A^S_i, i\in \NN)$  Bernoulli r.v. with parameter $p^-$;
\item $(A^N,A^N_i, i\in \NN)$  Bernoulli r.v. with parameter $p^+$;
\item $(B^S_i,B^S_{ji}, i,j\in \NN)$  Bernoulli r.v. with parameter
  $q^-$;
\item $(B^N_i,B^N_{ji}, i,j\in \NN)$  Bernoulli r.v. with parameter
  $q^+$.
\end{itemize}
Let
$D_i=1$ if agent $i$ is in state
$S$ and $D_i=0$ otherwise.
We define $\phi_i = D_iA^S_i +(1-D_i) A^N_i$.
The variable $\phi_i$  models the direct loss: if $\phi_i=1$ there is
a direct loss for agent $i$, otherwise there is no direct loss for
agent $i$. We also define $\theta_{ji} = D_i B^S_{ji} + (1-D_i) B^N_{ji}$.
The variable $\theta_{ji}$  models the possible contagion from
agent $j$ to agent $i$: if $\theta_{ji}=1$, there is contagion
otherwise there is no contagion.

Then the fundamental recursion satisfied by the vector $X=(X_i, i\in
V)$ is
\begin{eqnarray}
\label{eq:rec}1-X_i &=& (1-\phi_i)\prod_{j\sim i}(1-\theta_{ji}X_j).
\end{eqnarray}

\subsection{Epidemic risks for interconnected agents}

In order to completely specify our model, we still need to define how
to choose the variables $D_i$, i.e. whether agent $i$
invests in self-protection (corresponding to $D_i=1$) or not
($D_i=0$).

First,  note that the probability of loss for agent $i$ is  given,
depending on whether or not it invests in self protection, by
\begin{eqnarray}
p^S_i&:=&\EE[X_i|D_i=1]\label{eq:pS}
, \mbox{
  or, }\\
p^N_i&:=& \EE[X_i|D_i=0].
\label{eq:pN} 
\end{eqnarray}

In view of (\ref{eq:optic}), the best response of agent $i$ is given by:
\begin{eqnarray}
\label{eq:D}D_i=\ind(c_i < (p^N_i-p^S_i)\ell_i
+\pi_i[p^N_i]-\pi_i[p^S_i]),
\end{eqnarray}
where $p^S_i$ and $p^N_i$ are given by (\ref{eq:pS}) and
(\ref{eq:pN}).

Our model is defined by the graph $G$ (which topology is arbitrary)
and the set of Equations
(\ref{eq:rec},\ref{eq:pS},\ref{eq:pN},\ref{eq:D}).
In the rest of this paper,
we will make a simplifying assumption: we consider a heterogeneous
population, where agents differ only in self-protection cost and
potential loss.
The cost
of protection should not exceed the possible loss, hence $0\leq
c_i\leq \ell_i$.  The cost
$c_i$ and the potential loss $\ell_i$ are known to agent $i$ and varies among the population. Hence
we model this heterogeneous population by taking the sequence $(c_i,\ell_i
i\in \NN)$ as a sequence of i.i.d. random variables independent of
everything else.

So far, we have not yet specified the underlying graph.
We will consider random families of graphs $G^{(n)}$ with $n$
vertices and give asymptotic results as $n$ tends to infinity.
In all cases, we assume that the family of graphs $G^{(n)}$ is
independent of all other processes.

\section{Local Mean Field Model}\label{sec:lmf}

In this section, we introduce our Local Mean Field (LMF) model.
It extends the standard mean-field approximation by allowing to model
the correlation structure on local neighborhoods.
It can be shown that the LMF gives the exact asymptotic behavior of
the process $X$ as the number of vertices tends to infinity for sparse
random graphs with asymptotic given degree distribution $P(d)$ (see
\cite{durrett} for a definition).
A rigorous proof of this fact can be found in \cite{lelbol08} for a
particular case of the model described in Section \ref{sec:stepid}.
We will not attemp to give a general proof here. The main tool is the
notion of local weak convergence \cite{aldste}.

\subsection{Exact results for trees}

Since the graphs we are considering can be considered locally
to be like trees (with high probability), we first examine the
case where $G=T$ is a tree with nodes
$\o,1,\dots$ and a fixed root $\o$.

For a node $i$, we denote by
$\gen(i)\in \NN$
the generation of $i$, i.e. the length of the minimal path from $\o$
to $i$. Also we denote $i\to j$ if $i$ is a children of
$j$, i.e. $\gen(i)=\gen(j)+1$ and $j$ is on the minimal path from
$\o$ to $i$.
For an edge $(i,j)\in E$ with $i\to j$, we denote by $T_{i\to j}$ the sub-tree of
$T$ with root $i$ when deleting edge $(i,j)$ from $T$.
We have a family of trees $T_{i\to
  j}$ and we run the epidemic model according to equation
(\ref{eq:rec}) with the same variables
$(B^S_i,B^N_i,B_{ij}^S,B_{ij}^N,c_i, \ell_i ,i,j\in \NN)$ on each tree.
Hence the epidemics on the various subtree of $T$ are coupled thanks
to these random variables.
We say that node $i$ is infected from
$T_{i\to j}$ if the node $i$ is
infected in $T_{i\to j}$.
We denote by $Y_i$ the corresponding indicator function with value $1$
if $i$ is infected from $T_{i\to j}$ and $0$ otherwise. A simple
induction shows that the recursion (\ref{eq:rec}) becomes:
\begin{eqnarray}
\label{eq:rect}1-Y_i = (1-\phi_i) \prod_{k\to i}\left(
  1-\theta_{ki}Y_{k}\right).
\end{eqnarray}
If the tree $T$ is finite, we can compute all the $Y_i$ recursively starting
from the leaves with $Y_\ell=\phi_\ell$ for any leaf $\ell$.
As a consequence (and this is the main difference with
(\ref{eq:rec}) which makes the model on a tree tractable), the random
variables $Y_k$ with $k\to i$ in the right-hand term
of (\ref{eq:rect}) are independent of each others and independent of
the $\theta_{ki}$.
For any node $i\in T$, we just defined $Y_i$ and the
family $(Y_i, i\in T)$ is a tree-indexed process called a Recursive
Tree Process (RTP).

Consider now the case where $T$ is a Galton-Watson branching process
with offspring distribution $P^*$. The tree $T$ is now possibly infinite
but it is still possible to define an invariant RTP on $T$.
One way to construct it consists in defining a RTP for each finite
depth-$d$ tree and then show that these RTPs converge to an invariant
RTP as the depth $d$ tends to infinity \cite{aldban05}.
We first introduce the Recursive Distributional Equation (RDE):
\begin{eqnarray}
\label{eq:RDE} Y \=d 1-(1-\phi)\prod_{k=1}^{N^*} (1-\theta_k Y_k),
\end{eqnarray}
where $N^*$ has distribution $P^*$, $\phi=D A^S+(1-D) A^N$, $\theta_k = D
B^S_k+(1-D)B^N_k$ where $D$ is a Bernoulli r.v. with
parameter $\gamma $, $Y$ and $Y_k$
are i.i.d. copies.
We also assume that the random variables $D$, $A^S$, $A^N$, $B^S_k$,
$B^N_k$ and $Y_k$ are independent of each
others. Note however that $\phi$ and the $\theta$'s are not
independent of each others.
RDE for RTP plays a similar role as the equation $\mu=\mu K$ for the
stationary distribution of a Markov chain with kernel $K$, see
\cite{aldban05}.
The following result (proved in Appendix \ref{ap:RDE}) solves the RDE.
\begin{proposition}\label{prop:RDE}
For $p^+>0$, the RDE (\ref{eq:RDE}) has a unique
solution: $Y$ is a Bernoulli random variable with parameter
$h(\gamma)$, the unique solution in $[0,1]$ of
\begin{eqnarray*}
h = 1-\gamma (1-p^-) G_N(1-q^- h)-(1-\gamma) (1-p^+) G_N(1-q^+ h)
\end{eqnarray*}
where $G_{N^*}(x) =\EE[x^{N^*}]$ is the generating function of the
distribution $P^*$.
Moreover the function $\gamma \mapsto h(\gamma)$ is non-increasing in
$\gamma$.
\end{proposition}
As a consequence, we see that it is possible to construct an invariant
version of the RTP on the tree $T$ where for each $k\geq 0$, the sequence
$(Y_i, i\in T, \gen(i)=k)$ is a sequence of i.i.d. Bernoulli
random variables with parameter $h$, see \cite{aldban05}.

\subsection{LMF associated to a random network}
Our LMF model is characterized by the connectivity distribution $P(d)$
but the underlying tree $T$ as to be slightly modified compare to
previous section: if we start with a given vertex then the number of
neighbors (the first generation in the branching process) has
distribution $P$ but this is not true for the second generation.
Let $T$ be a Galton-Watson branching process with a root
which has offspring distribution $P$ and all other nodes have
offspring distribution $P^*$ given by $P^*(d-1) = \frac{dP(d)}{\sum
  dP(d)}$ for all $d\geq 1$.
\begin{remark}\label{rem:er}
Note that if $P$ is the Poisson distribution with parameter $\lambda$
which is the asymptotic degree distribution for Erdos-Renyi graph
$G(n,\lambda/n)$, then $P^*$ is also Poisson with mean $\lambda$.
\end{remark}
We now explain how to define the LMF based on the analysis made in
previous section.
Clearly, the crucial point in recursion (\ref{eq:rect}) is the fact
that the $Y_i$ can be computed ``bottom-up''. However a node can also
be infected from its parent and $Y_i$ is NOT a good approximation of
the real process $X_i$. Indeed the only node for which previous
analysis gives an approximation of the process $X$ is for the root and
the $Y_i$'s encode the information that the root is infected by an agent in
the subtree of $T$ ``below'' $i$.

Hence we define
\begin{eqnarray}
\label{eq:X}X(D)\=d 1-(1-\phi) \prod_{k=1}^{N}\left(1-\theta_{k}Y_{k}\right),
\end{eqnarray}
where $N$ has distribution $P$, $\phi$ and $\theta_k$ are the same as
in (\ref{eq:RDE}) and the $Y_k$'s are i.i.d. Bernoulli r.v. with
parameter $h(\gamma)$, i.e. satisfying
the RDE (\ref{eq:RDE}) with $N^*$ having distribution $P^*$.

\subsection{Asymptotic results}

We now show how to get quantitative results from our LMF.
The goal of Section \ref{sec:netext} is to derive such results for
various cases.

We consider a family of random graphs on $n$ vertices $G^{(n)}$ and
the associated process $(X^{(n)}_i, i\in
\{0,\dots,n-1\})$ satisfying the equations of our model on $G^{(n)}$.
We assume that our family of random graphs converges locally to a
tree as described in previsous section.
This property is true for sparse random graphs \cite{aldste}. It can
be shown that the process $X^{(n)}$ is asymptotically equivalent to
the process defined on the tree, i.e. the corresponding LMF model
described in previous section \cite{lelbol08}. Hence we restict our
analysis to the LMF model and the quantities computed here correspond
to the asymptotic values of the corresponding quantites for the
process $X^{(n)}$ for large values of $n$.

Let $\gamma$ be the fraction of the population investing in
self-protection. Then by symetry, the random variables $D_i$ are
i.i.d. Bernoulli r.v. with parameter $\gamma$. Thanks to the results
of the previous section, we can compute the law of the $X_i$'s. From this
law, we can compute the corresponding probability of loss depending on
the choice made to invest or not. Then one has to check self-consistency: the fraction of the population for which the best-response consists in investing in self-protection should be $\gamma$. Hence to solve our LMF model, we
need to solve the following fixed point equation:
\begin{eqnarray}
\nonumber p^{N,\gamma} &=& \EE[X(D)|D=0]\\
\label{f1}&=&1-\EE\left[ (1-A^N)\prod_{i=1}^N(1-B^N_iY_i)\right],\\
\nonumber p^{S,\gamma} &=& \EE[X(D)|D=1]\\
\label{f2}&=&1-\EE\left[ (1-A^S)\prod_{i=1}^N(1-B^S_iY_i)\right],\\
\label{f3}c^\gamma &=&
(p^{N,\gamma}-p^{S,\gamma})\ell+\pi[p^{N,\gamma}]-\pi[p^{S,\gamma}],\\
\label{f4}\gamma &=& \PP(c\leq c^\gamma),
\end{eqnarray}
where the distribution of $X(D)$ is given by (\ref{eq:X}) or
equivalently the $Y_i$ are i.i.d. Bernoulli r.v. with parameter
$h(\gamma)$ given by Proposition \ref{prop:RDE}.

Let $\gamma^*$ be a solution of this fixed point
equation. Then we have the following interpretations: $\gamma^*$ is
the fraction of the population investing in self-protection,
$p^{N,\gamma^*}$ is the probability of loss for an agent not investing
in self-protection and $p^{S,\gamma^*}$ is the probability of loss for
an agent investing in self-protection.
Hence the average probability of loss is
\begin{eqnarray*}
\EE[X(D)] = \gamma^* p^{S,\gamma^*}+(1-\gamma^*)p^{N,\gamma^*}.
\end{eqnarray*}

The outcome of rational behavior by
self-interested agents can be inferior to a centrally designed
outcome. By how much?
The price of anarchy, the most popular measure of
the inefficiency of equilibria, is defined as the ratio between the
worst objective function value of an equilibrium of the game and that
of an optimal outcome (possibly centralized in which case it will not
be described by the model introduced above).
In our setting, the cost incurred to agent $i$ is $c_i+
p^S_i\ell_i+\pi_i(p^S_i)$ if it invests in security and
$p^N_i\ell_i+\pi_i(p^N_i)$ otherwise. So for a given equilibrium, we
can compute the total cost incurred to the population. The price of
anarchy is the ratio of the largest (among all equilibria) such cost
divided by the optimal cost.
The price of anarchy is at least $1$ and a value close to $1$
indicates that the given outcome is approximately optimal. We refer to
\cite{agt} for an introduction to the inefficiency of equilibria (in
particular chapter 17).
We show in the next section how to compute this price of anarchy.

\section{Network externalities and the deployment of security
  features}\label{sec:netext}

We next use our LMF model to compare the following situations:
\begin{itemize}
\item Case 1: Strong protection. If an agent invest in
  self-protection, it cannot be harmed at all by the actions or
  inactions of others: $p^-=q^-=0$ (this is as in \cite{msw06})
\item Case 2: Weak protection. Investing in self-protection does not
  change the probability of contagion: $q^+=q^-$ (as in \cite{lelbol08})
\end{itemize}
In both cases, agents that invest in self-protection incur some cost
and in return receive some individual benefit through the reduced
individual expected loss. But part of the benefit is public, namely the
reduced indirect risk in the economy from which everybody benefits.
Hence, there is a negative externality associated with not
investing in self-protection, namely the increased risk to others.

\subsection{Erdos-Renyi graphs}

We analyze our model on a large sparse random graph
$G^{(n)}=G(n,\lambda/n)$ on $n$ nodes $\{0,1,\dots,n-1\}$, where each
potential edge $(i,j)$, $0\leq i<j\leq n-1$ is present in the graph
with probability $\lambda/n$, independently for all $n(n-1)/2$
edges. Here $\lambda>0$ is a fixed constant independent of $n$.
This corresponds to the case of the Erd\"os-R\'enyi graph which has
received considerable attention in the past \cite{durrett}. As
explained in Section \ref{sec:lmf}, our analysis is not restricted to
this class of graphs, but it is simpler in this case since
the degree distribution $P$ is a Poisson distribution with mean
$\lambda$ (see Remark \ref{rem:er}).

In this case, the fixed point equation for $h(\gamma)$ in Proposition
\ref{prop:RDE} becomes:
\begin{eqnarray*}
h = 1-\gamma (1-p^-) e^{-\lambda q^- h}-(1-\gamma) (1-p^+) e^{-\lambda
  q^+ h}.
\end{eqnarray*}
Then the equations (\ref{f1}) and (\ref{f2}) are given by:
\begin{eqnarray*}
p^{N,\gamma} &=& 1-(1-p^+)e^{-\lambda q^+ h(\gamma)},\\
p^{S,\gamma} &=& 1-(1-p^-)e^{-\lambda q^- h(\gamma)}.
\end{eqnarray*}
For simplicity, we drop the risk adverse condition, so that $\pi\equiv
0$ and we assume that costs for the self-protection are the same for
all agents and equal to $c$, and the possible losses are also the same and
equal to $\ell$.
Then we have
\begin{eqnarray*}
c^\gamma = \left((1-p^-)e^{-\lambda q^- h(\gamma)}-(1-p^+)e^{-\lambda q^+
  h(\gamma)}\right)\ell.
\end{eqnarray*}
Recall that an
agent decides to invest in self-protection iff $c<c^\gamma$.
The monotonicity of $c^\gamma$ in $\gamma$ is crucial and it depends
on the value of the parameters $(p^+,p^-,q^+,q^-)$.

\subsection{Case 1: Strong protection}\label{sec:fr}
We first consider  Case 1 where $p^-=q^-=0$, so that $p^{S,\gamma}=0$
and $c^\gamma = p^{N,\gamma} \ell = \left(1-(1-p^+)e^{-\lambda q^+
    h(\gamma)}\right)\ell$.
Then by Proposition \ref{prop:RDE}, $\gamma\mapsto c^\gamma$ is
non-increasing and the fixed point equation
(\ref{f1},\ref{f2},\ref{f3},\ref{f4}) has a unique solution.
\begin{figure}[htb]
\begin{center}
\includegraphics[width=4.5cm]{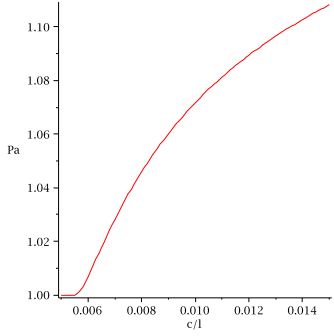}
\hspace{25pt} \caption{Price of anarchy for $c/\ell$ in the vicinity of $p^+=0.01$.} \label{fig:pa1}
\end{center}\end{figure}
In this case, as $\gamma$ the fraction of agents investing in self-protection
increases, the incentive to invest in self-protection decreases. In
fact, it is less attractive for an agent to invest in self-protection,
should others then decide to do so. As more agents invest, the
expected benefit of following suit decreases since there is a
reduction in the negative externalities which translates into a lower
probability of loss.
Hence there is a unique equilibrium point which is a Nash
equilibrium. However, there is a wide range of  parameters for which
the Nash equilibrium will not be socially optimal because agents do
not take into account the negative externalities they are creating in
determining whether to invest or not. Indeed it is easily shown that
at least for $c>p^+\ell$, the price of anarchy is strictly larger than
one (see Figure \ref{fig:pa1}).
\begin{proposition}
\label{prop:fr} The fixed point equation
(\ref{f1},\ref{f2},\ref{f3},\ref{f4}) reduces to
\begin{eqnarray*}
h = \frac{h\ell}{c}\left(1-(1-p^+)e^{-\lambda q^+h}\right) \mbox{ and, } 1-\gamma=\frac{h\ell}{c}.
\end{eqnarray*}
It has a unique solution.
The price of anarchy is given by
\begin{eqnarray*}
P_a(c) = \sup_{\gamma} \frac{c}{\gamma c+h(\gamma)\ell},
\end{eqnarray*}
where $h(\gamma)$ is the unique solution of
\begin{eqnarray*}
h=(1-\gamma)\left(1-(1-p^+)e^{-\lambda q^+h}\right)
\end{eqnarray*}
\end{proposition}
See Appendix \ref{ap:fr} for a proof.

\subsection{Case 2: Weak protection}
We now consider  Case 2 where $q^+=q^-$, so that $\gamma\mapsto
c^\gamma$ is non-decreasing.
The analysis of this case is described \cite{lelbol08} (see Proposition 5).
The situation is quite different from the results we derived for Case 1 above. In particular,
we can have two Nash equilibria involving everyone or no one
investing in security. When there are two Nash equilibria, the
socially optimal solution is always for everyone to invest: each agent
will find that the cost of investing in self-protection will be
justified if it does not incur any negative externalities and society
will be better off as well.
\begin{proposition}
We have $c^0<c^1$ and
\begin{itemize}
\item if $c<c^0$, then there is only one Nash equilibrium where every
  agent invest in self-protection;
\item if $c>c^1$, then there is only one Nash equilibrium where no
  agent invest in self-protection;
\item if $c^0<c<c^1$, then both Nash equilibria are possible.
\end{itemize}
The price of anarchy is given by:
\begin{eqnarray*}
P_a(c) = 1\vee \ind(c^0<c)\frac{h(0)\ell}{c+h(1)\ell}.
\end{eqnarray*}
\end{proposition}
If we take $p^-=0$, then we have $h(1)=0$ and $h(0)=h^*$ solution of
$h^* = 1-(1-p^+)e^{-\lambda q h^*}$. So that we have
\begin{eqnarray*}
P_a(c)\sim \frac{h^*\ell}{c}.
\end{eqnarray*}
\begin{figure}[htb]
\begin{center}
\includegraphics[width=4.5cm]{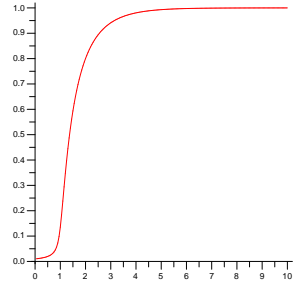}
\hspace{25pt} \caption{Price of anarchy: $h^*$ as a
  function of $\lambda q^+$, with $p^+=0.01$ and $p^-=0$.} \label{fig:price}
\end{center}\end{figure}
Figure \ref{fig:price} shows the value of $h^*$ as a function of
$\lambda q^+$. Note that typically $c=o(\ell)$ so that the price of
anarchy can be substantially larger than one.

\section{Discussion}
\label{sec:discussion}

We have shown that both situations with strong or weak protections
exhibit externalities and that the equilibrium is not socially optimal: therefore,
there is a market failure. However there are several important differences
to understand between strong and weak protections before trying to
resolve this market failure.

In case 1, the situation is similar to the free-rider problem which
arises in the production of public goods. If all then agents
invest in self-protection, then the general security level of the network
is very high since the probability of loss is  zero. But a
self-interested agent would not continue to pay for self-protection
since it incurs a cost $c$ for preventing only direct losses that have very
low probabilities. When the general security level of the network is
high, there is no incentive for investing in self-protection. This
results in an under-protected network.

Note that in this case, if the cost for self-protection is not
prohibitive, there is always a non-negligible fraction of the agents
investing in self-protection.
In case 2, the situation is quite different since no agent at all
invests in self-protection.
Even if a small fraction of agents does invest, and so raises the
general level of security of the network, it is not sufficient for the
benefit obtained by investing in self-protection for a new agent to be
larger than the cost of self-protection.

These facts seem very relevant to the situation observed in the
Internet, where under-investment in security solutions and
security controls has long been considered an issue. Security
managers typically face challenges in providing justification
for security investments, and in 2003, the President's National
Strategy to Secure Cyberspace stated that government action
is required where "market failures result in under-investment
in cybersecurity" \cite{whitehouse2003}.

It shows the power of our basic model to note that these interesting and very relevant phenomena emerge from
our analysis. Note also that these phenomena correspond to two extreme values of the parameter $q^-$,
namely case 1 corresponds to $q^-=0$ and case 2 corresponds to
$q^-=q^+$.
Hence taking $p^-=0$ and fixing all other parameters, we have a family
of models indexed by $q^-$, denoted simply $q$ in what follows, which
varies 'continuously' between the two cases.

Recall that $q$ is the probability of contagion when the agent invests
in self-protection. If $q=0$, the agent is completely secure whereas
for $q=q^-$, agents have the same probability of contagion whatever
their choices to invest or not in self-protection.
Hence $q$ can be interpreted as the inverse of the quality of the technology used for
self-protection.
\begin{figure}[htb]
\begin{center}
\includegraphics[width=5.5cm]{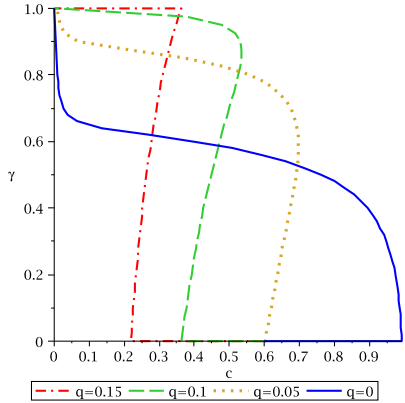}
\hspace{25pt} \caption{Adoption curves} \label{fig:adopt}
\end{center}\end{figure}
First note that when $q=0$, the technology is 'perfect' since there is
no possible loss. We are in the situation of case 1 and we see that
due to purely economic reasons, the technology is under-deployed in the
network because people 'free-ride' the benefit of the technology. Consider now
the case of an arbitrary $q$.
Figure \ref{fig:adopt} shows the adoption curves for different
values of $q$. This curve shows the fraction of the population investing in
security technology as a function of its cost (normalized by the
loss). Other parameters are $p^+=0.01, q^+=0.5$ and $\lambda =10$.

We observe some counter-intuitive phenomena.
First for a fixed price, increasing the quality of the security technology can
lead to a decrease of its adoption in the population!
Here is a qualitative interpretation of how this arises:
when the technology is not very good, propagation of the epidemic is
possible even if the agent uses the technology. Then agents have to pool their
efforts in order to compensate for the weakness of the technology. In
other words, a large number must invest in self-protection in order to
have an acceptable level of security.
But when the technology becomes better, then agents that did invest in
it start to step down from the group of investors and choose to
free-ride.

Second there is a barrier for choosing self-protection (except
  when $q=0$).  Namely for a fixed $q$, we see that there is a range
  for the parameter $c$ (close to $c^0$) such that the population is
  'trapped' in state $N$ whereas for the same values of the
  parameters, the situation where a large fraction of the population
  is investing would be a sustainable equilibrium point. There is a
  possibility of tipping or cascading: inducing some agents to invest
  in self-protection will lead others to follow suit. The curves of
  Figure \ref{fig:adopt} allow us to quantify the minimal number of
  agents to induce in order to trigger a large cascade of adoption.


\newpage
\section{Appendix}
\subsection{Proof of Proposition \ref{prop:RDE}}\label{ap:RDE}
Recall that the RDE is given by:
\begin{eqnarray*}
\label{eq:RDE} Y \=d 1-(1-\phi)\prod_{k=1}^{N^*} (1-\theta_k Y_k),
\end{eqnarray*}
where $N^*$ has distribution $P^*$, $\phi=D A^S+(1-D) A^N$, $\theta_k = D
B^S_k+(1-D)B^N_k$ where $D$ is a Bernoulli r.v. with
parameter $\gamma $, $Y$ and $Y_k$
are i.i.d. copies.
Let $h=\PP(Y=1)$, then we have
\begin{eqnarray*}
h&=& \PP\left(D=1, (1-A^S)\prod_{k=1}^{N^*}(1-B^S_kY_k)=0\right) \\
&&+\PP\left(D=0,
(1-A^N)\prod_{k=1}^{N^*}(1-B^N_kY_k)=0\right)\\
&=& \gamma(1-\PP(A^S=0))\EE\left[
  \PP(B^S_kY_k=0)^{N^*}\right]\\
&&+(1-\gamma)(1-\PP(A^N=0))\EE\left[
  \PP(B^N_kY_k=0)^{N^*}\right],
\end{eqnarray*}
and the first part of Proposition \ref{prop:RDE} follows.

We define:
\begin{eqnarray*}
f(x,\gamma) &=& 1 -\gamma(1-p^-)G_{N^*}(1-q^-x)\\
&&-(1-\gamma)(1-p^+)G_{N^*}(1-q^+x),
\end{eqnarray*}
so that $h$ is solution of the fixed point equation $h=f(h,\gamma)$.
By taking the derivate of $f$ in $x$, we see that $x\mapsto f(x,\gamma)$ is a
non-decreasing concave function.
Note that $f(0,\gamma)=\gamma p^-+(1-\gamma)p^+\geq (1-\gamma) p^+$
and $f(1,\gamma) \leq 1$. So that for $\gamma<1$, there exists a
unique solution to the fixed point equation $h=f(h,\gamma)$.
If $\gamma=1$, we have $f(0,1)=p^-$ and $f(1,1)<1$. Then if $p^-=0$,
the fixed point equation has a unique solution $h=0$ and if $p^->0$,
then $f(0,1)>0$ and the fixed point equation has still an unique
solution.

We now prove that the function $\gamma \mapsto h(\gamma)$ is
non-increasing. By taking the derivate of the function $\gamma\mapsto f(x,
\gamma)$, we see that this function is non-increasing in $\gamma$
(while $x$ is fixed).
Then for $u\leq v$, we get
\begin{eqnarray*}
f(h(u),u)=h(u) \geq f(h(u),v)\geq f(f(h(u),v),v)\geq h(v),
\end{eqnarray*}
and the claimed monotonicity of $h$ follows.

\subsection{Proof of Proposition \ref{prop:fr}}\label{ap:fr}

Recall that the fixed point equation for $h(\gamma)$ is:
\begin{eqnarray*}
h=(1-\gamma)\left( 1-(1-p^+)e^{-\lambda q^+h}\right).
\end{eqnarray*}
Consider now that the cost $c$ and loss $\ell$ are random variables
such that the function $t\mapsto \PP(c/\ell\leq t)$ is continuous,
then Equation (\ref{f4}) is
\begin{eqnarray*}
\gamma &=& \PP\left(c\leq \ell
  \left(1-(1-p^+)e^{-\lambda q^+ h(\gamma)}\right)\right)\\
&=& \PP\left(\frac{c}{\ell}\leq \frac{h(\gamma)}{1-\gamma}\right).
\end{eqnarray*}
Since the function $h$ is non-increasing, we see that the right-hand
side of the first line is a non-increasing function in $\gamma$, hence
there exists a unique solution $\gamma^*$ to this fixed point equation.
If we take a sequence of distributions such $c/\ell$ tends to a constant, we see
that the solution $\gamma^*$ is such that
\begin{eqnarray*}
\frac{c}{\ell} = \frac{h(\gamma^*)}{1-\gamma^*},
\end{eqnarray*}
and the first part of Proposition \ref{prop:fr} follows.

Note that we have $p^{N,\gamma}=h(\gamma)/(1-\gamma)$. So for a fixed
$\gamma$, the average cost incured to the population is
$\gamma c+(1-\gamma)p^{N,\gamma} \ell = \gamma c +h(\gamma)\ell$.
Now for $\gamma=\gamma^*$, we have $h(\gamma^*)\ell=(1-\gamma^*)c$,
so that the average cost is just $c$ and the last part of Proposition
\ref{prop:fr} follows.

\end{document}